\documentstyle[aps,prl,twocolumn,epsf]{revtex}
\begin{document}
\draft
\title{ Interaction effects on the conductance peak
height statistics in quantum dots}

\author{Y. Alhassid$^1$ and A. Wobst$^{1,2}$}

\address{$^1$Center for Theoretical Physics, Sloane Physics Laboratory,
              Yale University, New Haven, Connecticut 06520, USA\\
         $^2$Institut f\"ur Physik, Universit\"at Augsburg,
          86135 Augsburg, Germany}

\date {\today}

\maketitle
\begin{abstract}
 A random interaction matrix model is used to study
 the statistics of conductance peak heights in Coulomb blockade quantum
dots.  When the single-particle dynamics conserves time-reversal
 symmetry, the peak height statistics is insensitive to the
interaction strength. But when the single-particle dynamics
breaks time-reversal symmetry, the peak height statistics
exhibits a crossover from unitary to orthogonal
symmetry as the interaction strength increases. This crossover
 is driven by the time-reversal symmetry of the interaction. Our random
 interaction matrix model describes features of both the measured 
peak height and peak spacing statistics.
\end{abstract}


\narrowtext

  Random matrix theory (RMT) provides a useful tool for 
describing the universal statistical fluctuations of the spectrum and
 eigenfunctions of a quantum system whose associated classical
dynamics is chaotic. RMT has been successfully applied to the study of
mesoscopic phenomena in quantum dots -- submicron
2D devices where electrons are confined
by electrostatic potentials \cite{Kastner92}. In dots with
irregular shapes the single-electron dynamics is mostly chaotic, 
and the use of RMT is justified within a single-particle framework.
  In dots that are strongly coupled to leads, or open dots, the
 approximation of non-interacting
 quasi-particles is reasonable, and interactions can be
  considered indirectly through their effect on the electron
   coherence time. When a finite dephasing time is included, 
RMT can describe quantitatively the observed conductance
 statistics in open dots \cite{Huibers98}.

As the dot-leads coupling is made weaker, the
charge on the dot becomes quantized, and electron-electron
interactions cannot be ignored. In such almost closed dots, the
conductance displays sharp peaks versus the gate voltage. The 
measured peak height distributions \cite{Chang,Marcus} 
were found to agree
 with RMT predictions \cite{JSA92}. However, it is not clear
 why RMT should describe the peak height
statistics in dots with strong electron-electron interactions.
For example, the spacings between peaks were observed to have a
Gaussian-like distribution \cite{Sivan96,spacings} and not the
Wigner-Dyson distribution that is expected in the constant
interaction model plus single-particle RMT. Numerical
simulations of an Anderson model of a small disordered dot
 ($\lesssim 10$ electrons) with Coulomb interactions also
found Gaussian peak spacing distributions \cite{Sivan96},
while the peak height distributions showed only a weak dependence
 on interactions \cite{Berkovits98}.

 RMT is not limited to non-interacting systems. Indeed it
was first introduced to explain the neutron resonance data in
 the compound nucleus.  But the neutron resonances are measured
 at finite excitations, while the linear conductance experiments
 in quantum dots probe the ground state of the system with a varying
number of electrons. To understand the statistics that emerges
 from the interplay between one-body chaos and interactions
in these dots, it is necessary to construct a random matrix model
 that includes interactions
explicitly and reduces to one-body RMT in the absence of
interactions \cite{HAW98}.  We can break the time-reversal
 symmetry of the 
one-body Hamiltonian (e.g., with a magnetic field), but
 the time-reversal symmetry of the two-body interaction should
 be preserved.  An important question is whether such a random 
matrix model can reproduce both the measured peak height 
and peak spacing statistics.  

 A random interaction matrix model (RIMM) was introduced in
 Ref. \cite{RIMM} to study the peak spacing distribution in
 chaotic dots. The model combines a random two-body 
interaction \cite{French70} with a random one-body Hamiltonian,
 and describes a crossover of the peak spacing
statistics from a Wigner-Dyson distribution to a Gaussian
 distribution in terms of a parameter that measures the
 fluctuations of the interaction matrix elements.
 Here, we use the RIMM to study the effects of interactions
 on the conductance peak height statistics. 
 For a GOE one-body statistics,
 we find that the partial width (to decay into one of the
 leads) has a GOE Porter-Thomas
distribution independently of the interaction strength.
However, for a GUE one-body statistics, we find a crossover
from a GUE to a GOE Porter-Thomas distribution as the interaction
strength increases.  It is well known that, in the absence of
 interactions, a complete crossover from GOE to GUE statistics
 can be induced by a magnetic field. Our results indicate
 that this crossover is not complete when strong interactions 
are turned on because of a competition between an asymptotic
symmetry of the one-body Hamiltonian and the GOE symmetry
 of the time-reversal--conserving two-body interaction.  
 Finally, for the case of GUE
 one-body statistics, we find that a Gaussian-like peak
 spacing distribution develops at interaction strengths for which
the peak height statistics is still close to GUE.  This
explains the observed peak spacing statistics
\cite{Sivan96,spacings} and peak height statistics 
\cite{Chang,Marcus} within a single random matrix model.

  The $n$-th conductance peak height $G_n$ corresponds to the
tunneling of an electron into a dot with $n-1$ electrons to form
a dot with $n$ electrons.  At low temperatures,
$G_n =(e^2 /h)(\pi \bar{\Gamma}/{4 kT}) g_n$,
where \cite{Be91}
\begin{eqnarray}\label{peak}
   g_n =  {1 \over \bar{\Gamma}} \frac{\Gamma^l_n
\Gamma^r_n}{\Gamma^l_n + \Gamma^r_n} \;.
\end{eqnarray}
$\Gamma^{l(r)}_n$  is the partial width of the ground state of
the $n$-electron dot to decay into an electron in the left
(right) lead and the ground state of the dot with $n-1$ electrons:
\begin{equation}\label{width}
\Gamma_n \propto \left|\langle \Phi_{\rm g.s.}(n) |
 \;\psi^\dagger({\bbox r}) |\; \Phi_{\rm g.s.}(n-1) \rangle \right|^2 \;.
\end{equation}
$\psi^\dagger({\bbox r})$ is the creation operator of an
electron at the point $\bbox r$ [${\bbox r}={\bbox
r}_{l(r)}$ for the left (right) point contact], and $\Phi_{\rm
g.s.}(n)$ is the ground state wavefunction of the $n$-electron
dot. For non-interacting electrons, $\Phi_{\rm g.s.}(n)$ is a
Slater determinant of the $n$ lowest single-particle eigenfunctions
in the dot, and Eq.~(\ref{width}) reduces to
 $\Gamma_n \propto | \phi_n({\bbox r})|^2$, where
$\phi_n$ is the $n$-th single-particle wavefunction.
 If the
single-particle dynamics is chaotic, $| \phi_n({\bbox r})|^2$
satisfies Porter-Thomas statistics, leading to universal
distributions of the conductance peak heights (\ref{peak}) that
 are sensitive only to the underlying symmetry class \cite{JSA92}.

 To determine how interactions might
modify the Porter-Thomas statistics of the partial widths, we use
the RIMM of spinless interacting fermions \cite{RIMM}
\begin{eqnarray}\label{Hamiltonian}
H = \sum\limits_{ij} h_{ij} a^\dagger_i a_j +{1\over 4} \sum_{ijkl}
\bar u_{ijkl}a^\dagger_i a^\dagger_j a_l a_k
\;.
\end{eqnarray}
The one-body matrix elements $h_{ij}$ are chosen from the appropriate
 Gaussian random matrix ensemble, while the
 anti-symmetrized two-body matrix elements $\bar
u_{ij;kl}  = u_{ij;kl} - u_{ij;lk}$
form a GOE in the two-particle space \cite{French70}
\begin{eqnarray}\label{random-ensemble}
P(h) \propto e^{-{\beta\over 2a^2} {\rm Tr}\; h^2} \;;\;\;\; P(\bar u)
\propto   e^{- {\rm Tr}\;\bar u^2/2U^2}
\;.
\end{eqnarray}
 The states
$|i\rangle  = a_i^\dagger |0\rangle$ describe a fixed basis of $m$
single-particle states.  $h$ is an $m\times m$ GOE (GUE)
matrix when the single-particle dynamics conserves (breaks)
time-reversal symmetry.  The
two-body interaction is assumed to preserve time-reversal
symmetry and forms a GOE, irrespective of the symmetry of the
 one-body Hamiltonian.

   To calculate the statistics of the partial widths $\Gamma_n$,
 we expand $\psi^\dagger({\bbox r}) = \sum_{i} \psi_i({\bbox r})
a^\dagger_i$, where $\psi_i({\bbox r}) \equiv \langle {\bbox r}
| i \rangle$ is the wavefunction of the fixed state $|i\rangle$.
It follows from the orthogonal invariance of the ensemble
(\ref{random-ensemble}) that the statistics of $\Gamma_n$ is identical
 to the statistics of
\begin{equation}\label{width-i}
\Gamma^i_n \propto \left|\langle \Phi_{\rm g.s.}(n)
 | \; a^\dagger_i \; | \Phi_{\rm g.s.}(n-1) \rangle \right|^2
 \end{equation}
for any $i$. 

\begin{figure}
\epsfxsize=8.6 cm
\centerline{\epsffile{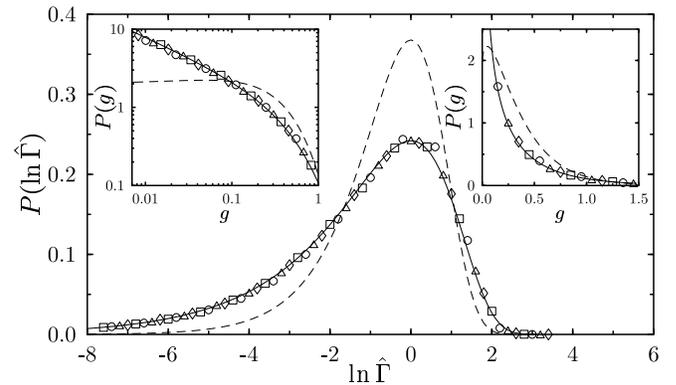}}
\vspace{3 mm}

\caption
{ Width and conductance peak height distributions in the RIMM
with GOE one-body statistics.  $P(\ln \hat \Gamma)$ us shown versus
 $\ln \hat \Gamma$ ($\hat \Gamma$ is the normalized width
 (\protect\ref{width-i})) for $m=12$, $n=4$ and $U/\Delta=0$
(circles), $2.4$ (squares), $4$ (diamonds) and $8$ (triangles). The
 solid and dashed lines are the GOE and GUE Porter-Thomas
distributions, respectively. Insets: the peak height
distributions $P(g)$ versus $g$ in a log-log scale (left inset) 
 and in a linear scale (right inset)  for
 the same cases shown in the main figure. The solid and dashed
lines are the GOE and GUE peak height distributions,
respectively.}
\label{fig1}
\end{figure}

For each realization $H$ of the ensemble (\ref{Hamiltonian}), we
calculate the ground states for $n-1$ and $n$ electrons, and compute
 $\Gamma_n^i$ using (\ref{width-i}). The
distributions of the normalized width  $\hat\Gamma = \Gamma/\bar
\Gamma$ for a GOE single-particle statistics are shown in
 Fig.~\ref{fig1} for several values of $U/\Delta$.
 We show $P(\ln \hat\Gamma)$ rather than $P(\hat \Gamma)$ in
order to display more clearly the small values of $\hat \Gamma$.
 The distributions are independent
of $U/\Delta$ and are well described by
the GOE Porter-Thomas distribution $P_{\rm GOE}(\ln \hat \Gamma) =
 (\hat\Gamma/2)^{1/2} \exp(-\hat\Gamma/2)$ (solid line).
 As a reference we also show the GUE
Porter-Thomas distribution $P_{\rm GUE}(\ln \hat\Gamma) = \hat\Gamma
\exp(-\hat\Gamma)$ (dashed line).  The conductance peak heights are
calculated from (\ref{peak}) using two uncorrelated ``sites'' $i$ and
$j$ for the left and right leads. The peak height
distributions $P(g)$ (for a GOE $h$) are shown in the insets of
Fig.~\ref{fig1}. They are all in good agreement with
 the GOE distribution $P_{\rm GOE}(g) =  \sqrt{{2 / \pi g}}
e^{-2 g}$ (solid lines) irrespective of $U/\Delta$.
 The dashed lines describe the GUE distribution
$P(g) = 4 g[K_0(2g) + K_1(2g)] \mbox{e}^{-2g}$ ($K_0$ and $K_1$
are Bessel functions).

\begin{figure}
\epsfxsize=8.6 cm
\centerline{\epsffile{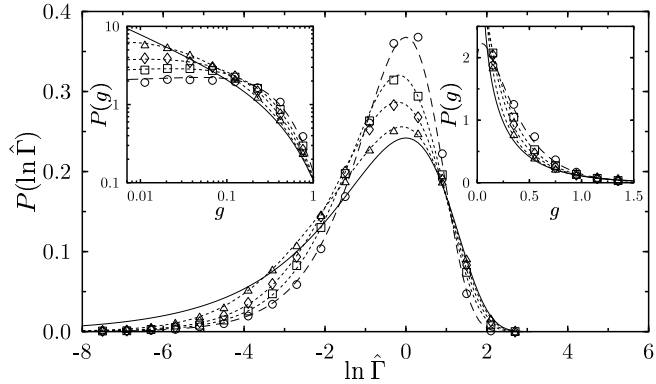}}
\vspace{3 mm}

\caption
{ Width and conductance peak height distributions in the RIMM
with GUE one-body statistics.  $P(\ln \hat \Gamma)$ is shown
 versus $\ln \hat \Gamma$ for the same cases as in
Fig.~\protect\ref{fig1}. The short-dashed lines are fits to the
theoretical distribution (\protect\ref{crossover}) describing the
RMT crossover from GUE ($\lambda=\infty$, dashed line) to
GOE ($\lambda=0$, solid line) with $\lambda=0.28$, $0.17$ and 
$0.08$, respectively.
Insets: $P(g)$ in
a log-log scale (left inset) and in a linear scale
(right inset) for the same cases shown in the main figure. The
short-dashed lines are the analytic peak height distributions in
the crossover regime \protect\cite{AHW98}
 between the GUE (dashed) and GOE (solid) distributions.
}
\label{fig2}
\end{figure}

 The results for a GUE one-body statistics are shown in
Fig.~\ref{fig2} for the same values of $U/\Delta$ as in Fig.~\ref{fig1}.
When the interaction strength increases, the width and peak height
distributions make a crossover from the corresponding GUE
distribution (at $U=0$) to the GOE distribution (at large $U$).
Equivalently, the transition from GOE to GUE statistics due to a
 time-reversal symmetry-breaking one-body field is not complete
 because of the competing GOE symmetry of the two-body interaction.  The
crossover distributions are compared with distributions
obtained from an RMT ensemble that describes the crossover
between the orthogonal and unitary symmetries. This ensemble is
 $H= S +i\alpha A$ , where $S$ and $A$ are $N\times N$ symmetric and
antisymmetric uncorrelated Gaussian matrices and $\alpha$ is a
parameter \cite{Mehta91}. Its wavefunction statistics 
depends on the parameter $\lambda \equiv \alpha \sqrt{N}/ \pi$. In
particular, the width distribution is given in closed
form by \cite{FE94,AHW98}
\begin{eqnarray}\label{crossover}
  P_\lambda(\hat{\Gamma}) = \int_0^1 {\rm d}t \; P_\lambda(t)
{ 1 +t^2 \over t} e^{-\left({1 + t^2 \over t}\right)^2 \hat{\Gamma}}
I_0 \left( { 1- t^4 \over t^2} \hat{\Gamma} \right)
\;,
\end{eqnarray}
where $I_0$ is the modified Bessel function of order zero.
$t$ in (\ref{crossover}) is a ``shape'' parameter that fluctuates
in the interval $[0,1]$ according to a known
 distribution \cite{FE94,AHW98,LBB97}
\begin{eqnarray}\label{t-dist}
 P_\lambda(t) = \pi^2 & & \left(1/t^3 -t \right) \lambda^2
 e^{-{\pi^2 \over 2} \lambda^2 \left(t- 1/t \right)^2}
 \left\{ \phi_1(\lambda) +  \right. \nonumber \\
& & \left. \left[ \left( t +
 t^{-1} \right)^2/4
 - 1 / (2\pi^2 \lambda^2) \right]  \left[ 1 - \phi_1(\lambda) \right]
\right\} \;,
\end{eqnarray}
where $\phi_1(\lambda)= \int_0^1  {\rm d}y \; e^{ - 2 \pi^2 \lambda^2
(1 - y^2)}$.  Eq.~(\ref{crossover}) reduces to the GOE and GUE
Porter-Thomas distributions for $\lambda=0$ and
$\lambda \gg 1$, respectively. In practice, the crossover from the
 GOE to the GUE already occurs for $\lambda \sim 1$.

 For each  $U/\Delta$, we find $\lambda$ by
 fitting the distribution (\ref{crossover}) to the
computed width distribution. The distributions (\ref{crossover}),
shown by the short-dashed lines in Fig.~\ref{fig2}, accurately
describe the width distributions of the RIMM with a GUE $h$.
 Good agreement with closed RMT expressions  \cite{AHW98}
 is also obtained for the peak height distributions $P_\lambda(g)$
shown in the insets of Fig.~\ref{fig2}. 

\begin{figure}
\epsfxsize=8.6 cm
\centerline{\epsffile{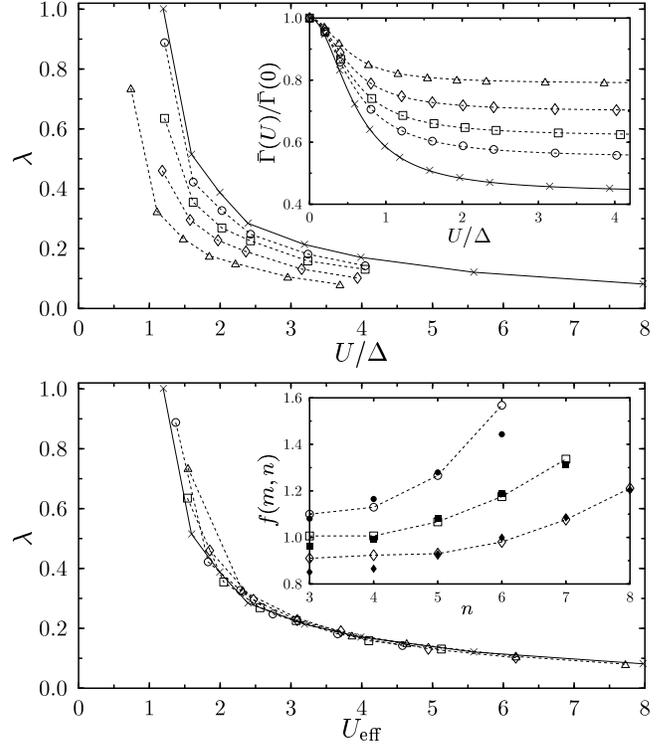}}
\vspace{3 mm}

\caption
{ Top: The RMT crossover parameter $\lambda$ as a function of
$U/\Delta$ for $m=10$ and $n=4$ (circles), $5$ (squares), $6$
(diamonds), and $7$ (triangles). The solid line is the
reference curve $m=12$, $n=4$.
 Top inset: The average width $\bar \Gamma(U)/\bar
\Gamma(0)$ versus $U/\Delta$ for the same cases shown in
the main figure. The lines are fit to $\bar\Gamma(U)
 - \bar\Gamma(\infty) = (\bar\Gamma(0)-\bar\Gamma(\infty))/
 (1 + b\, U^2/\Delta^2)$.
 Bottom: The curves from the top
panel scale on the reference curve (solid line)
 when plotted versus $U_{\rm eff} \equiv f(m,n)\, U/\Delta$.
  The bottom inset shows
 the scaling factor $f(m,n)$ (open symbols) versus $n$ for
 $m=10$ (circles), $12$ (squares), and $14$ (diamonds).
  The scaling factors are nearly the same as the ones
 derived from the peak spacing statistics \protect\cite{RIMM}
(solid symbols). }
\label{fig3}
\end{figure}

   An important issue is the universality associated with the
RIMM (\ref{Hamiltonian}) and
(\ref{random-ensemble}).  The model depends on three parameters:
 $m$, 
 $n$, and $U/\Delta$. The top panel of
Fig.~\ref{fig3} shows, for a GUE one-body $h$, the crossover
parameter $\lambda$ as a function of $U/\Delta$ for $m=10$ and
$n=4,5,6,7$ (symbols), and for a ``reference'' case  $m=12$ and
$n=4$ (solid line). The curves depend
on both $m$ and $n$, but can all be scaled on the reference
 curve after scaling the interaction
strength by a constant, $U_{\rm eff} \equiv f(m,n)\, U/\Delta$ 
(see bottom panel of Fig.~\ref{fig3}).
  The values of the scaling factor $f(m,n)$, shown in the bottom
 inset of Fig. ~\ref{fig3}, are essentially the same as those needed to
obtain a universal peak spacing statistics \cite{RIMM}. We
conclude that the peak spacing statistics as well as the
 peak height statistics in the RIMM depend only on a
 single parameter $U_{\rm eff}$.

 In the range $U_{\rm eff} \lesssim 1$, $\lambda \geq 1$ and the
  statistics is essentially GUE.
 For $U_{\rm eff} \sim 1 - 1.5$, the 
 peak height statistics is close to GUE while the peak spacings
 already follow a Gaussian distribution. This explains 
the measured RMT-like peak height distributions
 \cite{Chang,Marcus}  and the Gaussian-like shape of
 the peak spacing distributions 
\cite{Sivan96,spacings} within a
 single random matrix model. We remark that the small
deviations from GUE statistics observed in the experiment of Ref.
 \cite{Chang} and in the calculations of Ref. \cite{Berkovits98}
 in the presence of a magentic field
are consistent with a crossover from GUE to GOE.

  The average width $\bar \Gamma$ is a monotonically
 decreasing function of $U/\Delta$ and saturates at large values
of $U$. The top inset of Fig.~\ref{fig3} shows $\bar
\Gamma(U)/\bar \Gamma(0)$  as a function of $U/\Delta$ for
several $m$ and $n$. This dependence is well described by
$\bar\Gamma(U) - \bar\Gamma(\infty) =
(\bar\Gamma(0)-\bar\Gamma(\infty))/(1 + b\,U^2/\Delta^2)$. The
parameter $\bar\Gamma(\infty)$ depends on $m$ and $n$, while $b$
is found to be independent of $m$ and to depend only weakly on
$n$.

Next we compare the predictions of the RIMM with those of a
model of a quantum dot. We studied
a 2D Anderson model with
on-site disorder parameter $W$ and hopping matrix element $V=1$.
The electrons are interacting with a Coulomb interaction whose
strength over one lattice spacing $a$ is $U_c=e^2/a$
 \cite{Berkovits98}.   We choose periodic boundary conditions
 in both directions so that the average width and width 
statistics are independent of the specific
sites to which the leads are attached.  With these boundary 
conditions, a background charge (that was included in the 
calculations of Ref. \cite{Berkovits98}) only shifts the 
energy levels by an $n$-dependent constant but does not 
affect the wavefunctions.

\begin{figure}
\epsfxsize=8.6 cm
\centerline{\epsffile{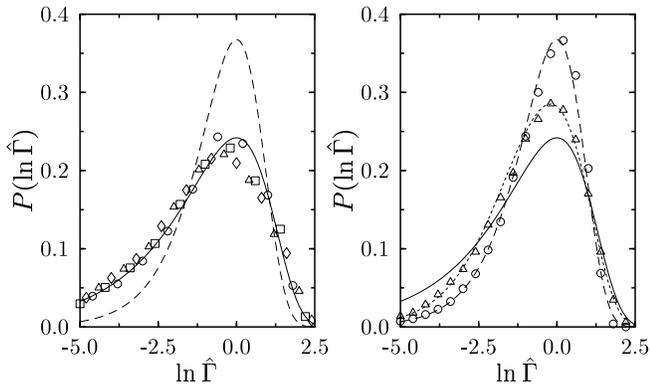}}
\vspace{3 mm}

\caption
{Width distributions in a $4\times 5$ Anderson model plus Coulomb
interactions.
 Left: $P(\ln \hat \Gamma)$  in the absence
of a magnetic field and for several values of the Coulomb interaction
strength: $U_c=0$ (circles), $2$ (squares), $4$ (diamonds) and
 $6$ (triangles). The disorder parameter is $W=5$. The solid
 (dashed) line is the GOE (GUE)
Porter-Thomas distribution.  Right: $P(\ln \hat \Gamma)$  for $U_c=0$
 (circles) and $12$ (triangles) in the presence of time-reversal
symmetry-breaking magnetic flux $\Phi=0.14 \Phi_0$ and for a
disorder strength of $W=3$.  The short-dashed line is the
distribution (\ref{crossover}) with $\lambda=0.17$.
}
\label{fig4}
\end{figure}

  The left panel of Fig.~\ref{fig4} shows the width statistics in the
absence of magnetic flux  for a $4\times 5$ lattice with
 disorder parameter $W=5$ and for $n=4$ electrons. The
distributions are approximately described by the GOE 
Porter-Thomas distribution (solid line) irrespective of the value
 of $U_c$ and in agreement with the RIMM.  The dependence of the average
 width $\bar \Gamma$ on the interaction strength (not shown)
 is similar to that observed in the RIMM. The right panel shows width
distributions in the presence of magnetic flux $\Phi =0.14 \Phi_0$
and for $W=3$. The $U_c=0$ distribution (circles) agrees with the
 GUE Porter-Thomas distribution (dashed line), while the $U_c=12$
 distribution (triangles) is described by
(\ref{crossover}) with $\lambda=0.17$. This is the distribution
obtained in the RIMM for $U_{\rm eff} \approx 4$.
For weaker interactions the calculated width distributions
exhibit some deviations from (\ref{crossover}).
 We note however that, for the small
lattices used, it is difficult to find a disorder
strength for which the model is in the metallic diffusive regime
and displays universal RMT statistics. In particular, in the presence
 of a magnetic flux we could not find values of $W$ for which both
 the spectral and wavefunction statistics are GUE.

In conclusion, we have investigated the width and peak height
 statistics in Coulomb-blockade quantum dots using a random
interaction matrix model with interactions that preserve 
time-reversal symmetry. For a GOE one-body symmetry the
 statistics is insensitive to the interaction strength. 
However, at strong interactions, a time-reversal 
symmetry-breaking field leads only  to a partial crossover 
from GOE to GUE statistics.  Our random interaction
 matrix model can reproduce both the observed Gaussian-like
 shape of the peak spacing distribution and the RMT statistics
 of the peak height distributions. 

We thank Y. Gefen for useful discussions. This work was 
supported in part by the U.S. DOE
grant No.\ DE-FG-0291-ER-40608. A.W. acknowledges a
  fellowship from the
Studienstiftung des Deutschen
Volkes.

\end{document}